\documentclass[letterpaper,english,prl,notitlepage,twocolumn]{revtex4-1}
\usepackage[T1]{fontenc}
\usepackage[latin9]{inputenc}
\usepackage{amsmath}
\usepackage{amssymb}
\usepackage{esint}

\makeatletter

\pdfpageheight\paperheight
\pdfpagewidth\paperwidth

\usepackage{color}
\usepackage[usenames,dvipsnames,svgnames,table]{xcolor}
\usepackage{graphicx}
\usepackage{comment}
\newcommand{\vect}[1]{\boldsymbol{\mathbf{#1}}}
\definecolor{violet}{rgb}{0.58, 0.0, 0.83}
\newcommand{\gae}{\lower 2pt \hbox{$\,
\buildrel{\scriptstyle >}\over {\scriptstyle \sim}\,$}}
\newcommand{\lae}{\lower 2pt \hbox{$\,
\buildrel{\scriptstyle <}\over {\scriptstyle \sim}\,$}}
\newcommand{\Tr}{\mathrm{Tr}}

\makeatother

\usepackage{babel}
\begin{document}

\title{Measuring second Chern number from non-adiabatic effects}

\author{
Michael Kolodrubetz}
\affiliation{Department of Physics, Boston University, 590 Commonwealth Ave., Boston, MA 02215, USA} \affiliation{Department of Physics, University of California, Berkeley, CA 94720, USA}
\affiliation{Materials Sciences Division, Lawrence Berkeley National Laboratory, Berkeley, CA 94720, USA}
\begin{abstract}
The geometry and topology of quantum systems have deep connections
to quantum dynamics. In this paper, I show how to measure the non-Abelian
Berry curvature and its related topological invariant, the second
Chern number, using dynamical techniques. The second Chern number
is the defining topological characteristic of the four-dimensional
generalization of the quantum Hall effect and has relevance in systems
from three-dimensional topological insulators to Yang-Mills field
theory. I illustrate its measurement using the simple example of a
spin-$3/2$ particle in an electric quadrupole field. I show how one
can dynamically measure diagonal components of the Berry curvature
in an over-complete basis of the degenerate ground state space and
use this to extract the full non-Abelian Berry curvature. I also show
that one can accomplish the same ideas by stochastically averaging
over random initial states in the degenerate ground state manifold.
Finally I show how this system can be manufactured and the topological
invariant measured in a variety of realistic systems, from superconducting
qubits to trapped ions and cold atoms.
\end{abstract}
\maketitle

Topological invariants such as the first Chern number have become
relevant in condensed matter physics due to their robustness in describing
novel states of matter \citep{Thouless1982_1,Niu1985_1,Kane2005_1,Fu2007_1,Fu2007_2}.
While naturally defined in the solid state Brillouin zone, these geometry
concepts and the Berry phase on which they are based occur in a wide
variety of systems. In particularly, these ideas have been recently
applied to engineer and measure topological properties of designed
systems, such as many-body cold atomic systems \citep{Miyake2013_1,Aidelsburger2013_1,Aidelsburger2015_1}
and few-body systems of qubits or random walkers \citep{Kitagawa2011_1,Schroer2014_1,Roushan2014_1}.

It was noted in the early days of topological physics \citep{Avron1988_1,Avron1989_1}
that higher topological invariants could be defined, and particularly
that a non-trivial second Chern number characterizes systems with
time-reversal symmetry. More recently, this has been connected the
four-dimensional generalization of the quantum Hall effect \citep{Zhang2001_1}
and three-dimensional topological insulators \citep{Essin2009_1}.
It is also intricately related to the axion electrodynamics used to
define 3D topological insulators and to non-perturbative instanton
effects in Yang-Mills field theory.

This higher topological invariant has never been measured experimentally.
Here I propose how the second Chern number may be measured using non-adiabatic
effects similar to the methods used in Refs. \citep{Schroer2014_1}
and \citep{Roushan2014_1} to measure the first Chern number. The
proposal relies on time-reversal invariant Hamiltonians to enforce
a doubly-degenerate ground state and thus the previous proposal must
be extended to account for these degeneracies. This involves measuring
a fundamentally non-Abelian topological object. We show two ways to
account for this - one by deterministically sampling over degenerate
ground states and another by stochastic sampling. Each method has
its pluses and minuses that may be relevant for different experimental
systems, and we close by discussing how to access this physics in
current experiments.

\begin{figure}[b]
\includegraphics[width=\columnwidth]{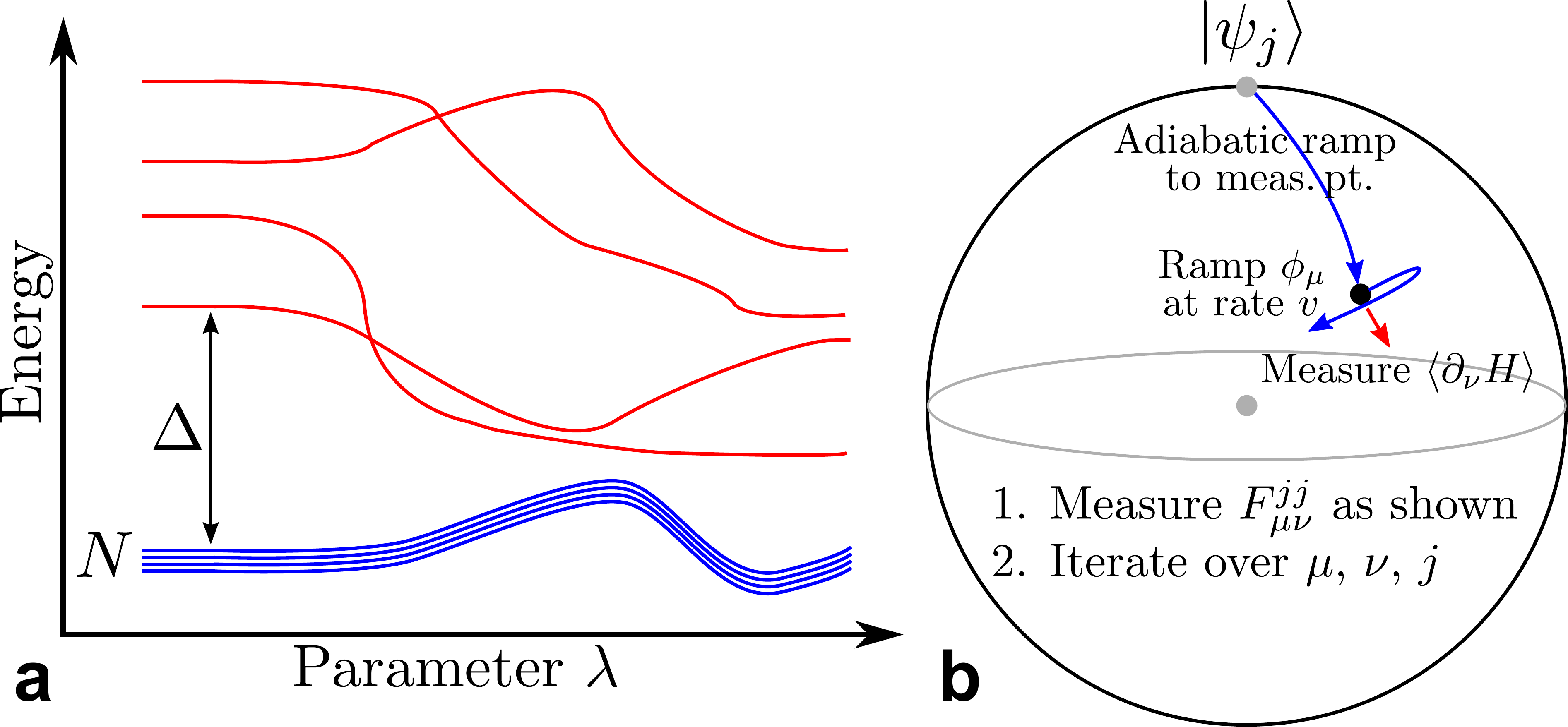}
\caption{Measuring second Chern number dynamically. (a) General setup where measurement is possible. $N$ degenerate ground states are separated by a non-zero gap from the excited states. The ground states must remain degenerate and gapped from the excited states, but no there are no restrictions on the excited states. (b) Illustration of the ramping protocol to find one component $F_{\mu \nu}^{jj}$ of the non-Abelian Berry curvature at measurement point $\vect \phi$. The procedure must be iterated over $\mu,\nu \in \{1,2,3,4\}$, and $j \in \{1,2,\ldots,N^2\}$ at each measurement point.}
\label{fig:measure_second_chern}
\end{figure}

\emph{Non-adiabatic corrections with degeneracies - }Consider a Hamiltonian
$H(\vect{\lambda})$ that depends on some parameters $\vect{\lambda}$.
If one starts in the non-degenerate ground state $|\psi_{0}(\vect{\lambda}_{i})\rangle$
at $\vect{\lambda}_{i}$ and then ramps $\vect{\lambda}$ slowly with
time, at zeroth order the system simply remains in it ground state
and picks up both a dynamical and Berry phase. If the ground state
remains non-degenerate during the course of the ramp, the leading
non-adiabatic correction leading to population in the excited states
can be calculated through a technique known as adiabatic perturbation
theory \citep{Rigolin2008_1,DeGrandi2010_2}. This can be understood
by translating the problem to a moving frame $|\psi\rangle\to U(\vect{\lambda})^{\dagger}|\psi\rangle$,
where $U(\vect{\lambda})$ diagonalizes the Hamiltonian ($H_{d}=U^{\dagger}HU$).
In the moving frame, the Hamiltonian becomes 
\begin{equation}
H_{m}=H_{d}-i\dot{\lambda}_{\mu}U^{\dagger}\partial_{\mu}U\equiv H_{d}-\dot{\lambda}_{\mu}\mathcal{A}_{\mu}^{m},\label{eq:H_moving_frame}
\end{equation}
where $\partial_{\mu}\equiv\partial/\partial\lambda_{\mu}$ and repeated
indices are summed over. We refer to the second term $\mathcal{A_{\mu}}=U\mathcal{A}_{\mu}^{m}U^{\dagger}$
as the Berry connection operator as its matrix elements in the energy
eigenbasis $|\psi_{n}\rangle$ are $\langle\psi_{m}|\mathcal{A}_{\mu}|\psi_{n}\rangle=i\langle\psi_{m}(\vect{\lambda})|\partial_{\mu}\psi_{n}(\vect{\lambda})\rangle$.
Diagonal terms in $H_{m}$ give rise to the dynamical and Berry phases,
since $A_{\mu}=\langle\psi_{0}|\mathcal{A}_{\mu}|\psi_{0}\rangle$
is the ground state Berry connection. Off-diagonal terms in this operator
give rise to non-adiabatic occupation in the excited states, which
at leading order can be seen by applying static perturbation theory
to $H_{m}$ to find $c_{n\neq0}\approx\dot{\lambda}_{\mu}\langle\psi_{n}|\mathcal{A}_{\mu}|\psi_{0}\rangle/(E_{n}-E_{0})$.
Calculating the ``generalized force'' $M_{\nu}\equiv-\langle\partial_{\nu}H\rangle$
in this state gives a leading order correction proportional to the
ground state Berry curvature, $M_{\nu}\approx M_{\nu}^{0}-\dot{\lambda}_{\mu}F_{\mu\nu}$,
where $F_{\mu\nu}=\partial_{\mu}A_{\nu}-\partial_{\nu}A_{\mu}$ \citep{Gritsev2012_1}.
This term is analogous to the ``anomalous velocity'' that appears
in the semi-classical theory of Bloch electrons and can be thought
of as a Lorentz force in parameter space. It has been used in few-qubit
experiments to measure the (Abelian) Berry curvature with non-degenerate
ground state manifolds, from which the topologically-invariant first
Chern number can be extracted \citep{Schroer2014_1,Roushan2014_1}.

This formalism has been generalized to situations where the ground
state is degenerate in a series of papers by Rigolin and Ortiz \citep{Rigolin2010_1,Rigolin2012_1,Rigolin2014}.
For the simplest case where the ground state remains $N$-fold degenerate
throughout the process, we can reformulate aspects of these results
rather simply through the above formalism. First note that, unlike
the non-degenerate case, the connection and curvature are now non-Abelian,
meaning they give rise not only to phases but more generally rotations
within the ground state subspace. In our language of adiabatic perturbation
theory in the moving frame, these non-Abelian effects can be seen
as first-order degenerate perturbation theory; at a given point $\vect{\lambda}$
during the ramp, one must diagonalize $\mathcal{A}_{\mu}^{m}$ within
the degenerate subspace, the eigenstates of which then just pick up
separate Berry phases as in the non-degenerate case. The non-Abelian
aspect comes as the diagonal basis of $\mathcal{A}_{\mu}^{m}$ changes
with $\vect{\lambda}$. From integrating Eq. \ref{eq:H_moving_frame},
we see that the anholonomy is given up to a dynamical phase by the
path-ordered integral $\mathcal{P}\mathrm{exp}\left[i\int_{\vect\lambda_{i}}^{\vect\lambda}d\lambda_{\mu}^{\prime}A_{\mu}(\vect{\lambda^{\prime}})\right]$.
For the degenerate case, where $A_{\mu}$ is now an $N\times N$ matrix
with matrix elements $A_{\mu}^{ij}=i\langle\psi_{0i}|\partial_{\mu}\psi_{0j}\rangle$
giving the non-Abelian Berry connection within the ground state sector.\footnote{More accurately, this derivation holds if the path-ordered integral
and the matrix $A_{\mu}$are represented in the moving frame by just
integrating the moving-frame Schrodinger equation within the degenerate
subspace (the upper $N\times N$ block). Care must be taken in defining
these anholonomies for large paths in parameter space, as a non-trivial
$C_{2}$ serves as an obstruction to defining a global $U(N)$ gauge.}

Fortunately, the off-diagonal terms responsible for excitations do
not notice this degeneracy. To see this, consider a path $\vect\lambda(s)$
such that an adiabatic traversal would yield the state $|\psi_{0A}(\vect\lambda)\rangle$
within the ground state sector. Tracing the same path at a finite
rate, the ground state component of the wave function is unchanged
at order $\dot{\lambda}$. Excitations do occur at this order, given
by the natural extension of the earlier formula:
\[
|\psi(\vect\lambda(t))\rangle\approx|\psi_{0A}(\vect\lambda)\rangle+i\dot{\lambda}_{\mu}\sum_{n\neq0}|\psi_{n}(\vect\lambda)\rangle\frac{\langle\psi_{n}|\partial_{\mu}\psi_{0A}\rangle}{E_{n}-E_{0}}.
\]
One may readily confirm that the generalized force in this state is
simply related to the diagonal component of the non-Abelian Berry
curvature matrix $F_{\mu\nu}=\partial_{\mu}A_{\nu}-\partial_{\nu}A_{\mu}-i[A_{\mu},A_{\nu}]$,
namely $M_{\nu}\approx M_{\nu A}^{0}-\dot{\lambda}_{\mu}F_{\mu\nu}^{AA}$.
Note that the adiabatic value $M_{\nu A}^{0}$ depends on the state
$|\psi_{0A}\rangle$. Thus our results for the physical observable
are clearly similar to the non-degenerate case, but with the important
caveat that they depend on the history of the protocol. This is because
$|\psi_{0A}\rangle$ depends on the path taken, so two different paths
that give the same value of $\vect\lambda$ and $\dot{\vect\lambda}$
at time $t$ will not necessarily give the same Berry curvature correction
to the generalized force.

\emph{Measuring second Chern number - }The question then becomes what
to make of the non-Abelian Berry curvature measurement if one can
not easily predict the adiabatically-connected state. We are left
searching for quantities that are invariant to the choice of basis.
We find such quantities in the topologically invariant Chern numbers.
The simplest example is the first Chern number, defined for a closed
two-dimensional manifold $\mathcal{M}_{2}$ in parameter space as
$C_{1}=(2\pi)^{-1}\int_{\mathcal{M}_{2}}d\lambda_{\mu}\wedge d\lambda_{\nu}\Tr(F_{\mu\nu})$,
where $\wedge$ denotes the wedge product . A novel topological invariant
that appears for the four-dimensional manifold $\mathcal{M}_{4}$
is the second Chern number

\begin{eqnarray}
C_{2} & = & \int_{\mathcal{M}_{4}}\omega_{2}^{\mu\nu\rho\sigma}d\lambda_{\mu}\wedge d\lambda_{\nu}\wedge d\lambda_{\rho}\wedge d\lambda_{\sigma}\label{eq:second_chern}\\
\omega_{2}^{\mu\nu\rho\sigma} & = & \frac{\Tr(F_{\mu\nu}F_{\rho\sigma})-\Tr(F_{\mu\nu})\Tr(F_{\rho\sigma})}{32\pi^{2}},\nonumber 
\end{eqnarray}
where and $\omega_{2}$ is the second Chern form. The trace is taken
over the ground state (upper) indices, i.e., $\Tr(F_{\mu\nu}F_{\rho\sigma})\equiv F_{\mu\nu}^{ij}F_{\rho\sigma}^{ji}$,
rendering $C_{2}$ basis invariant. But one clearly requires knowledge
of the off-diagonal elements of $F$ to take this trace, while our
non-adiabatic scheme only yields diagonal elements. I will now discuss
two schemes to fill in this gap, which may be suitable for different
systems.

First, let us see how we can deterministically reconstruct the matrix
$F$ by measuring its diagonal elements in an over-complete basis.
For concreteness, assume there are $N=2$ degenerate ground states,
denoted $|\psi_{0A}\rangle$ and $|\psi_{0B}\rangle$. $F_{\mu\nu}^{ij}$
is anti-symmetric w.r.t. exchange of the lower (parameter) indices
and Hermitian w.r.t. the upper (ground state) ones. Thus each matrix
$F_{\mu\nu}$ is determined by $N^{2}$ real numbers. If we measure
the diagonal components in the four states $|\psi_{1}\rangle=|\psi_{0A}\rangle$,
$|\psi_{2}\rangle=|\psi_{0B}\rangle$, $|\psi_{3}\rangle=(|\psi_{0A}\rangle+|\psi_{0B}\rangle)/\sqrt{2}$,
and $|\psi_{4}\rangle=(|\psi_{0A}\rangle+i|\psi_{0B}\rangle)/\sqrt{2}$,
then with a bit of algebra
\begin{eqnarray}
F_{\mu\nu} & = & \left(\begin{array}{cc}
F_{\mu\nu}^{AA} & F_{\mu\nu}^{AB}\\
F_{\mu\nu}^{BA} & F_{\mu\nu}^{BB}
\end{array}\right)\nonumber \\
 & = & \left(\begin{array}{cc}
F_{\mu\nu}^{11} & \frac{2iF_{\mu\nu}^{33}+2F_{\mu\nu}^{44}-(1+i)(F_{\mu\nu}^{11}+F_{\mu\nu}^{22})}{2i}\\
(F_{\mu\nu}^{AB})^{\ast} & F_{\mu\nu}^{22}
\end{array}\right),\label{eq:F_munu_deterministic}
\end{eqnarray}
from which evaluating the second Chern number integral is just math.
This method is well-suited to controllable quantum systems such as
qubits, ions, or ultracold atoms where one has the ability to prepare
arbitrary initial states. It trivially generalizes to arbitrary $N$.

If one does not have such a degree of control, a similar result may
be achieved stochastically. The central idea is that the object $\langle\psi|F_{\mu\nu}|\psi\rangle\langle\psi|F_{\rho\sigma}|\psi\rangle$
averaged over states $|\psi\rangle$ drawn uniformly from the ground
state subspace contains information about the second Chern form. In
particular, one can show that if $\overline{(~)}$ denotes this state
average, then
\begin{eqnarray}
\overline{\langle\psi|F_{\mu\nu}|\psi\rangle\langle\psi|F_{\rho\sigma}|\psi\rangle} & = & \frac{\Tr(F_{\mu\nu}F_{\rho\sigma})+\Tr(F_{\mu\nu})\Tr(F_{\rho\sigma})}{N(N+1)}\nonumber \\
\overline{\langle\psi|F_{\mu\nu}|\psi\rangle} & = & \frac{\Tr(F_{\mu\nu})}{2N}.\label{eq:stochastic_avg_general}
\end{eqnarray}
This can be readily seen for $N=2$, for which $\langle\psi|F_{\mu\nu}|\psi\rangle\langle\psi|F_{\rho\sigma}|\psi\rangle=[\cos^{2}(\theta/2)F_{\mu\nu}^{AA}+\cos(\theta/2)\sin(\theta/2)(e^{i\phi}F_{\mu\nu}^{AB}+c.c.)+\sin^{2}(\theta/2)F_{\mu\nu}^{BB}][\mu\nu\to\rho\sigma]$.
Averaging over the Bloch angles $\theta$ and $\phi$, phases $e^{in\phi}$
vanish and one readily reproduces Eq. \ref{eq:stochastic_avg_general}.
The general formula is derived in Appendix \ref{app:stochastic_chern_form}.
This method of measurement is natural if instead of deterministically
preparing the desired states, nature gives one access to random snapshots
of the system but allows multiple non-destructive measurements of
the same state, such that all components $F_{12}$, $F_{13}$, etc.
may be measured. This may therefore be more natural in the solid state
context.

\emph{Spin-3/2 in electric quadrupole field -} I now demonstrate the
applicability of the above measurement techniques on the quintessential
example of a system with non-trivial second Chern number: the quantum
spin-3/2 in an electric quadrupole field. This model was proposed
by Avron et al. \citep{Avron1988_1,Avron1989_1} as containing the
simplest ``quaternionic singularity'' in much the same way that
the monopole singularity of the Berry curvature in a qubit yields
a non-trivial first Chern number. The Hamiltonian may be written as
$H=-\vect{\lambda}\cdot\vect{H}$, where $\vect H=(H_{0},H_{1},\ldots,H_{4})$
denotes an orthonormal basis of spin-3/2 quadrupole operators and
$\vect\lambda$ denotes vector of coupling parameters. In particular,
we choose the basis described in Ref. \citealp{Avron1989_1}: $H_{0}=(-J_{x}^{2}-J_{y}^{2}+2J_{z}^{2})/3$,
$H_{1}=(J_{x}J_{z}+J_{z}J_{x})/\sqrt{3}$, $H_{2}=(J_{y}J_{z}+J_{z}J_{y})/\sqrt{3}$,
$H_{3}=(J_{x}^{2}-J_{y}^{2})/\sqrt{3}$, and $H_{4}=(J_{x}J_{y}+J_{y}J_{x})/\sqrt{3}$.
These Hamiltonians are invariant under time reversal, thus the eigenvalues
come in two degenerate pairs. By construction, the energy eigenvalues
of each are $\pm1$; due to orthonormality, this is also true for
arbitrary unit 5-vector $\vect\lambda$. 

It is clear from the above discussion that the only way for all four
eigenvalues to be degenerate is to have $\vect\lambda=\vect0$; this
is the ``quaternionic monopole'' that gives a non-zero second Chern
number. Avron et al. showed that for a 4-sphere surrounding this degeneracy,
the second Chern number is equal to 1. Furthermore, due to time reversal
symmetry this system has a vanishing first Chern number, so $C_{2}$
is its defining topological invariant. I will now show how the above
ideas can be used to measure $C_{2}$ directly, focusing on the deterministic
method for concreteness.

Let us begin by fixing the magnitude $|\vect\lambda|=1$ and re-parameterizing
the problem in terms of the spherical angles $\vect\phi=(\phi_{1},\phi_{2},\phi_{3},\phi_{4})$,
where $\phi_{4}\in[0,2\pi)$, $\phi_{1-3}\in[0,\pi)$, $\lambda_{0}=\cos\phi_{1}$,
$\lambda_{1}=\sin\phi_{1}\cos\phi_{2}$, $\ldots$, $\lambda_{4}=\sin\phi_{1}\sin\phi_{2}\sin\phi_{3}\sin\phi_{4}$.
To obtain the Chern form at some point $\vect\phi$, we begin with
one of the states $|\psi_{1-4}\rangle$ described earlier for the
value $\vect\phi=\vect0$ (the North pole). Here the Hamiltonian has
the simple form $H=5/4-J_{z}^{2}$, so that ground states are just
the $m_{z}=\pm3/2$ eigenstates. Starting from one of these states,
say $|\psi_{1}\rangle$, we ramp slowly along some arbitrary path
$\vect\phi_{1}(s)$ to the measurement point $\vect\phi$. Then to
measure the component $F_{\mu\nu}^{11}$, we ramp the parameter $\phi_{\mu}$
according to $\phi_{\mu}=\phi_{\mu}^{m}+v(t-t_{m})t^{2}/t_{m}^{2}$,
where $\phi_{\mu}^{m}$ is its value at the point to be measured.
This ramp is chosen such that the ramp starts smoothly ($\dot{\phi}_{\mu}(0)=0$)
at $\phi_{\mu}(0)=\phi_{\mu}^{m}$ and returns to $\phi_{\mu}^{m}$
at time $t_{m}$ with velocity $v$. Repeating this ramp multiple
times with velocity $v$ and $v/2$ and measuring the expectation
values $M_{\nu}(v)$ and $M_{\nu}(v/2)$, the Berry curvature is $F_{\mu\nu}^{11}\approx2[M_{\nu}(v/2)-M_{\nu}(v)]/v$.
This protocol, illustrated in Fig. \ref{fig:measure_second_chern},
must be repeated for all pairs $(\mu,\nu)$ and all initial states
$|\psi_{i}\rangle$ to obtain the second Chern form at the point $\vect\phi$
via Eq. \ref{eq:F_munu_deterministic}. Crucially, for a given point
$\vect\phi$, the same path $\vect\phi_{1}$ must be taken for each
component of the tensor to ensure that the appropriate phase relations
between the $|\psi_{i}\rangle$'s remain once ramped to $\vect\phi$.
From the second Chern form, the second Chern number may be obtained
by the integral in Eq. \ref{eq:second_chern}. Such higher-dimensional
integrals are numerically tricky; here I do it by Monte Carlo sampling
$\vect\phi$ uniformly from its domain. Carrying out the above procedure
yields $C_{2}=0.9926\pm0.0073$, consistent with the exact value of
1. 

\begin{figure}
\includegraphics[width=.8\columnwidth]{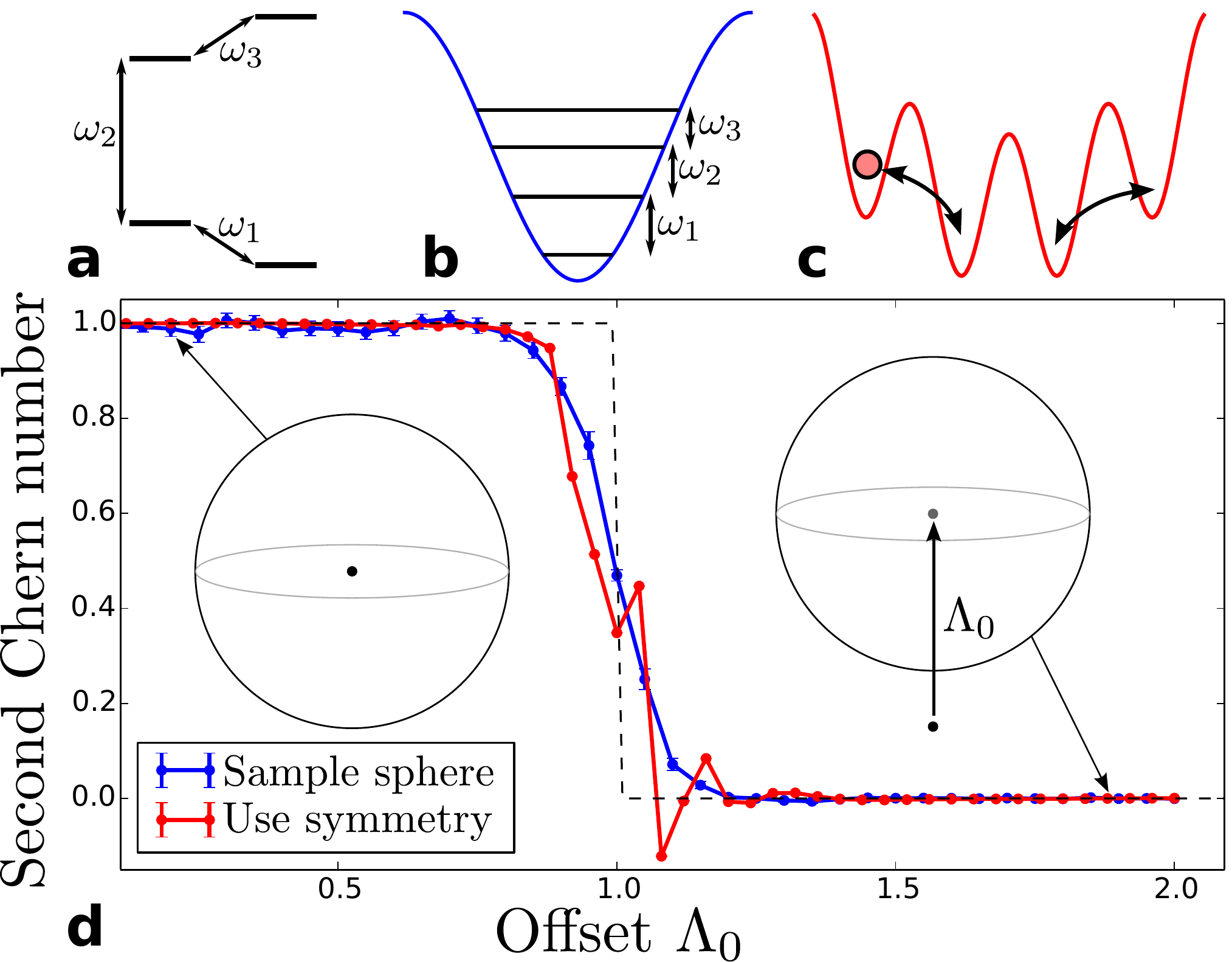} \caption{Experimental realizations of four-level systems where the second Chern number may be measured: (a) atomic hyperfine levels, (b) bound states of an artificial atom, (c) a particle hopping in a 4-site lattice. While all sites/levels must be coupled to realize arbitrary $4\times 4$ Hamiltonians, only the indicated drives/hoppings are necessary in the presence of symmetry about the $\lambda_0$ axis. (d) Second Chern number measured dynamically for the model of a spin-$3/2$ in an electric quadrupole field, as described in the text, either without utilizing symmetry (blue points) or with symmetry (red points). The deviation from quantization near the transition at $\Lambda_0=1$ is due to using finite time protocols: I ramp from the North pole to the measurement point in time $t=100$ then ramp for measurement with parameters $v=0.01$ and $t_m=100$.  The inset shows how as the offset term $\Lambda_0 H_0$ is added, the 4-sphere shifts away from the origin until eventually the (4-fold) degeneracy at $\vect \lambda = \vect 0$ is no longer enclosed, causing a topological transition at $\Lambda_0=1$.}
\label{fig:second_chern_expts}
\end{figure}

To demonstrate the robustness of this topological invariant, we may
induce a topological transition by adding a constant offset $\Lambda_{0}H_{0}$
to the previous Hamiltonian. This shifts the unit sphere by an amount
$\Lambda_{0}$, and for $|\Lambda_{0}|>1$ the sphere fails to surround
the degeneracy at the origin. Therefore, the second Chern number jumps
to being trivial. This topological transition is seen in the simulations
in Fig. \ref{fig:second_chern_expts}d; the transition appears broadened
for a finite velocity $v$ due to higher-order non-adiabatic corrections
near the gapless transition point.

\emph{Experiments - }The above procedure naturally lends itself to
controllable quantum systems such as superconducting qubits, ultracold
atoms, ions, and solid state defects. For such systems, more detailed
topological and geometric properties such as the Wilson loop may be
measured via full tomography of adiabatic protocols; the dynamic second
Chern number measurement serves to supplement this natural list of
tools. However, the dynamical measurement trivially generalizes to
more complicated systems where full tomography is not possible, requiring
neither strict adiabaticity nor tomographic measurements that scale
exponentially with system size.

An important practical concern in realizing these ideas experimentally
is to use the symmetry of the sphere to reduce the number of measurements
that must be made. As the simplest example of this, consider the case
$\Lambda_{0}=0$ where the problem has full spherical symmetry. Then,
since all points on the sphere are identical, the Chern number can
be obtained by measuring the Berry curvature at a single point. If
we start in the ground state at the North pole as before, then there
are four orthonormal tangent vectors to the surface: $\hat{\lambda}_{1}$,
$\hat{\lambda}_{2}$, $\hat{\lambda}_{3}$, and $\hat{\lambda}_{4}$.
We can measure the response to these parameters in the same way as
we did with $\vect\phi$'s. For instance, we obtain $F_{\lambda_{\mu}\lambda_{\nu}}$
by ramping $\lambda_{\mu}$ then measuring $M_{\lambda_{\nu}}=-\langle H_{\nu}\rangle$.
By symmetry, $\Tr(F_{\lambda_{1}\lambda_{2}}F_{\lambda_{3}\lambda_{4}}-F_{\lambda_{1}\lambda_{3}}F_{\lambda_{2}\lambda_{4}}+F_{\lambda_{1}\lambda_{4}}F_{\lambda_{2}\lambda_{3}})=3\Tr(F_{\lambda_{1}\lambda_{2}}F_{\lambda_{3}\lambda_{4}})$
and furthermore, this value will be constant in the basis of tangent
vectors at any point on the sphere. So we simply multiply by the surface
area of the unit 4-sphere to get\footnote{Note that because of time-reversal symmetry, the first Chern form
vanishes: $\Tr(F_{\mu\nu})=0$. Therefore the second Chern form reduces
to $\omega_{2}^{\mu\nu\rho\sigma}=\Tr(F_{\mu\nu}F_{\rho\sigma})/32\pi^{2}$.}
\begin{equation}
C_{2}=\frac{3A_{S^{4}}\Tr(F_{\lambda_{1}\lambda_{2}}F_{\lambda_{3}\lambda_{4}})}{4\pi^{2}}=2\Tr(F_{\lambda_{1}\lambda_{2}}F_{\lambda_{3}\lambda_{4}}).\label{eq:C2_sph_symm}
\end{equation}
We thus expect that $\Tr(F_{\lambda_{1}\lambda_{2}}F_{\lambda_{3}\lambda_{4}})=1/2$,
which is readily confirmed numerically.

The argument must be slightly modified in the presence of an offset
$\Lambda_{0}$, but symmetry still significantly reduces the number
of measurements required. $\lambda_{0}$ is now distinct from the
other axes, which translates into a $\phi_{1}$-dependence of the
second Chern form. The axes tangent to a point $\vect\phi=(\phi_{1},0,0,0)$
are now $\hat{\phi}_{1}$, $\hat{\lambda}_{2}$, $\hat{\lambda}_{3}$,
and $\hat{\lambda}_{4}$, so the non-trivial terms $F_{\phi_{1}\lambda_{2}}$
and $F_{\lambda_{3}\lambda_{4}}$ are obtained by ramping $\phi_{1}$
and $\lambda_{3}$ and measuring $\langle H_{2}\rangle$ and $\langle H_{4}\rangle$
respectively. For a given value of $\phi_{1}$, the remaining parameters
trace out a 3-sphere of radius $\sin\phi_{1}$, which has surface
area $A_{S^{3}}=2\pi^{2}\sin^{3}\phi_{1}$. By the same logic as Eq.
\ref{eq:C2_sph_symm}, one may then compute $C_{2}$ via
\begin{eqnarray}
C_{2} & = & \frac{3}{4\pi^{2}}\int_{0}^{\pi}d\phi_{1}A_{S^{3}}(\phi_{1})\Tr\left[F_{\phi_{1}\lambda_{2}}(\phi_{1})F_{\lambda_{3}\lambda_{4}}(\phi_{1})\right]\nonumber \\
 & = & \frac{3}{2}\int_{0}^{\pi}d\phi_{1}\sin^{3}\phi_{1}\Tr\left[F_{\phi_{1}\lambda_{2}}(\phi_{1})F_{\lambda_{3}\lambda_{4}}(\phi_{1})\right].\label{eq:C2_cyl_symm}
\end{eqnarray}
The resulting Chern number is shown in Fig. \ref{fig:second_chern_expts}d.

In addition to reducing the number of measurements, Eq. \ref{eq:C2_cyl_symm}
reduces the number of control axes required; one need only ramp $\phi_{1}$(i.e.
$\lambda_{1}$ and $\lambda_{2}$) and $\lambda_{3}$, although given
the symmetry of the problem any three $\lambda$s will do. This is
reduced further to only two $\lambda$s in the fully symmetric case.
While one may in principle realize arbitrary $4\times4$ Hamiltonians
given four levels full-connected by drives \footnote{Note that the matrix elements of the arbitrary $4\times4$ Hamiltonians
may be controlled by tuning the amplitude, phase, and detuning of
the drives.}, a more natural situation is partially-connected levels like the
ladder system illustrated in Fig. \ref{fig:second_chern_expts} \citep{Stuhl2015_1,Mancini2015_1,Wall2016_1}.
For such couplings, not all of the terms can be easily realized, but
fortunately a sufficient number can be realized to allow the measurement
using symmetry. This can be seen from the representations of $H_{i}$
in the $J_{z}$ basis: 
\[
H_{0}=\left(\begin{array}{cccc}
1 & 0 & 0 & 0\\
0 & -1 & 0 & 0\\
0 & 0 & -1 & 0\\
0 & 0 & 0 & 1
\end{array}\right),\:H_{1}=\left(\begin{array}{cccc}
0 & 1 & 0 & 0\\
1 & 0 & 0 & 0\\
0 & 0 & 0 & -1\\
0 & 0 & -1 & 0
\end{array}\right),\,\mathrm{etc.}
\]
If we think of the four states as realizing a single particle in a
4-site chain, then $H_{0}$ represents on-site chemical potentials,
while $H_{1}$ and $H_{2}$ represent (phased) hopping between sites
1 and 2 as well as 3 and 4. This is well within the capacity of the
driven system, and even could be realized on a physical 4-site lattice
or 4-site supercell within a larger lattice via lattice shaking schemes
analogous to those used to generate artificial gauge fields \citep{Miyake2013_1,Aidelsburger2013_1}.

Instead of imprinting the Hamiltonian structure on Hilbert space by
hand, we might instead be given a system with natural structure of
its own, say a set of coupled spins-1/2 \citep{Childress2006_1,Koch2007_1,Kim2010_1}.
In this case, there are two representations that may be useful. First,
note that we can write the Hamiltonians in the form $H_{0}=\sigma_{1}^{z}\sigma_{2}^{z}$,
$H_{1}=\sigma_{1}^{x}\sigma_{2}^{z}$, $H_{2}=\sigma_{1}^{y}\sigma_{2}^{z}$,
$H_{3}=\sigma_{2}^{x}$, and $H_{4}=\sigma_{2}^{y}$. Of these operators,
$H_{0}$, $H_{3}$ and $H_{4}$ are naturally realized for chains
of two spin-1/2's, in both trapped ions \citep{Kim2010_1} and transmon
qubits \citep{Strauch2003}. Similarly, we can imagine directly obtaining
$J=3/2$ by fusing three spin-1/2's. In this language, $H_{0}$ is
proportional to an Ising-like interaction $\sum_{<ij>}\sigma_{i}^{z}\sigma_{j}^{z}$
in a fully-couple three-spin ring. The other interactions seem less
natural: $H_{1}$ maps to $\sum_{<ij>}\sigma_{i}^{x}\sigma_{j}^{z}$,
similarly for $H_{2}$ and $H_{4}$, while $H_{3}$ gives interactions
of the form $\sigma_{i}^{x}\sigma_{j}^{x}-\sigma_{i}^{y}\sigma_{j}^{y}$.
With developments in the field, this three-qubit realization may be
possible in the future, while the two-qubit version is readily available
with current technology.

\emph{Discussion - }Most of this paper has worked in the language
of controllable quantum systems, for which it is natural to discuss
some set of control parameters $\vect\lambda$. In solid state physics,
a natural parameter space is the momenta $\vect k$ in the Brillouin
zone, or more generally twists of the boundary condition \citep{Thouless1983_1,Niu1984_1}.
For four-dimensional crystals, the insulator analogous to our spin-3/2
example is the second Chern insulator, which is the natural four-dimensional
generalization of the quantum Hall effect \citep{Zhang2001_1}; recent
work has proposed realizing this with cold atoms in an artificial
4D lattice \citep{Price2015_1}. One obtains a quantized non-linear
electromagnetic response in these systems. An interesting open question
is how to relate this to our method for determining the second Chern
number, which is obtained through \emph{linear} response, following
by classical post-processing. The stochastic method for extracting
the second Chern form bears greater resemblance to the 4D Hall response,
and is seemingly more natural in the solid state context. However,
it remains a product of linear responses, and its relationship to
non-linear fluctuations remains unclear at present.

\emph{Acknowledgments - }I would like to acknowledge useful conversations
with Claudio Chamon, Joel Moore, Anatoli Polkovnikov, Ana-Maria Rey,
Seiji Sugawa, and Jun Ye. During preparation of the manuscript, I
became aware of independent experimental work by the Spielman group
to measure the second Chern number via related techniques\cite{SugawaInPrep2016_1}.
I am pleased to acknowledge support from AFOSR FA9550-13-1-0039 as
well as Laboratory Directed Research and Development (LDRD) funding
from Berkeley Lab, provided by the Director, Office of Science, of
the U.S. Department of Energy under Contract No. DEAC02-05CH11231.

\onecolumngrid
\appendix

\section{Derivation of stochastic Chern form\label{app:stochastic_chern_form}}

In this appendix, we will derive Eq. \ref{eq:stochastic_avg_general}
for generic $N$. We start by noting that an $N\times N$ Hilbert
space may be uniformly sampled via picking a random unit vector $\hat{n}$
from $S^{2N}$ (the $2N$-sphere) and constructing the state $|\hat{n}\rangle=(n_{0}+in_{1})|0\rangle+(n_{2}+in_{3})|1\rangle+\cdots+(n_{2N-2}+in_{2N-1})|N-1\rangle\equiv\sum_{j=0}^{N-1}\eta_{j}|j\rangle$.
For an arbitrary Hermitian operator $A$, the expectation value in
$|\hat{n}\rangle$ is
\[
\langle\hat{n}|A|\hat{n}\rangle=|\eta_{0}|^{2}A_{00}+\eta_{0}^{\ast}\eta_{1}A_{01}+\ldots
\]
Averaging over the $2N$-sphere, only the terms without phase factors
survive. Then $\overline{|\eta_{0}|^{2}}=\overline{|\eta_{j}|^{2}}=\overline{n_{0}^{2}+n_{1}^{2}}=2\overline{n_{0}^{2}}$.
Expressing $\hat{n}$ in spherical coordinates $(\phi_{1},\phi_{2},\ldots,\phi_{2N-1})$,
where $\phi_{2N-1}\in[0,2\pi)$ and $\phi_{j\neq2N-1}\in[0,\pi)$,
we have $n_{0}=\cos\phi_{1}$, $n_{1}=\sin\phi_{1}\cos\phi_{2}$,
and area element $\mathrm{d}a=\sin^{2N-2}\phi_{1}\cdots\sin\phi_{2N-2}\mathrm{d}\phi_{1}\cdots\mathrm{d}\phi_{2N-1}$.
Thus
\[
\overline{n_{0}^{2}}=\frac{\int\cos^{2}\phi_{1}\mathrm{d}a}{\int\mathrm{d}a}=\frac{\int_{0}^{\pi}\cos^{2}\phi_{1}(\sin\phi_{1})^{2N-2}d\phi_{1}}{\int_{0}^{\pi}(\sin\phi_{1})^{2N-2}d\phi_{1}}=\frac{1}{2N}.
\]
Similarly, averaging the quantity $\langle\hat{n}|A|\hat{n}\rangle\langle\hat{n}|B|\hat{n}\rangle$
for $A$ and $B$ Hermitian gives
\begin{eqnarray}
\overline{\langle\hat{n}|A|\hat{n}\rangle\langle\hat{n}|B|\hat{n}\rangle} & = & \overline{|\eta_{0}|^{4}}A_{00}B_{00}+\overline{|\eta_{1}|^{4}}A_{11}B_{11}+\ldots+\overline{|\eta_{0}|^{2}|\eta_{1}|^{2}}(A_{00}B_{11}+A_{11}B_{00}+A_{01}B_{10}+A_{10}B_{01})+\ldots\nonumber \\
 & = & \overline{|\eta_{0}|^{4}}(A_{00}B_{00}+A_{11}B_{11}+\ldots)+\nonumber \\
 &  & \overline{|\eta_{0}|^{2}|\eta_{1}|^{2}}(A_{00}B_{11}+A_{11}B_{00}+A_{01}B_{10}+A_{10}B_{01}+A_{00}B_{22}+\ldots),\label{eq:avg_AB_basic}
\end{eqnarray}
where as before we have used the symmetry of the sphere to replace
everything by $\eta_{0}$ and $\eta_{1}$. Then
\begin{eqnarray*}
\overline{|\eta_{0}|^{4}} & = & \overline{(n_{0}^{2}+n_{1}^{2})^{2}}=\overline{n_{0}^{4}}+2\overline{n_{0}^{2}n_{1}^{2}}+\overline{n_{1}^{4}}=2\left(\overline{n_{0}^{4}}+\overline{n_{0}^{2}n_{1}^{2}}\right)\\
\overline{|\eta_{0}|^{2}|\eta_{1}|^{2}} & = & \overline{(n_{0}^{2}+n_{1}^{2})(n_{2}^{2}+n_{3}^{2})}=4\overline{n_{0}^{2}n_{1}^{2}}\\
\overline{n_{0}^{4}} & = & \frac{\int\cos^{4}\phi_{1}\mathrm{d}a}{\int\mathrm{d}a}=\frac{3}{4N(N+1)}\\
\overline{n_{0}^{2}n_{1}^{2}} & = & \frac{\int\cos^{2}\phi_{1}\sin^{2}\phi_{1}\cos^{2}\phi_{2}\mathrm{d}a}{\int\mathrm{d}a}=\frac{1}{4N(N+1)}.
\end{eqnarray*}
Inserting these results into Eq. \ref{eq:avg_AB_basic}, we find
\begin{eqnarray*}
\overline{\langle\hat{n}|A|\hat{n}\rangle\langle\hat{n}|B|\hat{n}\rangle} & = & \frac{2}{N(N+1)}(A_{00}B_{00}+A_{11}B_{11}+\ldots)+\frac{1}{N(N+1)}(A_{00}B_{11}+A_{11}B_{00}+A_{01}B_{10}+A_{10}B_{01}+A_{00}B_{22}+\ldots)\\
 & = & \frac{1}{N(N+1)}\Big[(A_{00}B_{00}+A_{11}B_{11}+\ldots+A_{01}B_{10}+A_{10}B_{01}+\ldots)+\\
 &  & \;\;\;\;\;\;\;\;\;\;\;\;\;\;\;\;\;\;(A_{00}B_{00}+A_{11}B_{11}+\ldots+A_{00}B_{11}+A_{11}B_{00}+\ldots)\Big]\\
 & = & \frac{1}{N(N+1)}\left[\Tr(AB)+(\Tr A)(\Tr B)\right],
\end{eqnarray*}
which clearly reduces to Eq. \ref{eq:stochastic_avg_general} when
$A=F_{\mu\nu}$ and $B=F_{\rho\sigma}$.

\bibliographystyle{aipnum4-1}
\bibliography{measure_second_chern_arxiv}

\end{document}